\documentclass[prb,twocolumn,superscriptaddress, showpacs,amsmath,amssymb]{revtex4}
\usepackage{graphicx}
\usepackage{subfigure}

\begin{document}

\title{Applications of electrostatic capacitance and charging}

\author{Titus Sandu}
\affiliation{National Institute for Research and Development in Microtechnologies-IMT,
126A, Erou Iancu Nicolae Street, 077190, Bucharest, ROMANIA}
\email{titus.sandu@imt.ro}
\author{George Boldeiu}
\affiliation{National Institute for Research and Development in Microtechnologies-IMT,
126A, Erou Iancu Nicolae street, 077190, Bucharest, ROMANIA}
\author{Victor Moagar-Poladian}
\affiliation{National Institute for Research and Development in Microtechnologies-IMT,
126A, Erou Iancu Nicolae street, 077190, Bucharest, ROMANIA}

\date{\today}

\begin{abstract}
The capacitance of an arbitrarily shaped object is calculated with the same second-kind integral equation method used for computing static and 
dynamic polarizabilities. The capacitance is simply the dielectric permittivity multiplied by the area of the object and divided by 
the squared norm of the Neumann-Poincar\'{e} operator eigenfunction 
corresponding to the largest eigenvalue. The norm of this eigenfunction varies slowly with shape thus enabling the definition of 
two scale-invariant shape factors and perturbative calculations of capacitance. The result is extended to a special class of capacitors in 
which the electrodes are the equipotential surfaces generated by the equilibrium charge on 
the object. This extention allows analytical expressions of capacitance for confocal spheroidal capacitors and finite cylinders. Moreover, a second order 
formula for thin constant-thickness capacitors is given with direct applications for capacitance of membranes in living cells 
and of supercapacitors. For axisymmetric geometries a fast and accurate numerical method is provided. 

\end{abstract}

\maketitle

\section{Introduction}
Potential theory has been proved very successfully in solving 
some 
boundary value problems such as the Dirichlet and the Neumann 
problems or the electrostatic charge distribution on conductors. For domains 
with sufficiently smooth boundaries (i.e., a regular piecewise Lyapunov surface) 
the above problems use specific types of 
potentials like the volume, the single-, and the double-layer potentials, 
the logarithmic potential for two-dimensional domains, etc. \cite{Kellog1967,Khavinson2007}. 
The Dirichlet and Neumann problems defined on domains with sufficiently 
smooth boundaries can be recast in integral equations which lead to compact operators on domain 
boundary: the Neumann-Poincar\'{e} (or double-layer) operator and its adjoint \cite{Khavinson2007}. These 
methods are applied in some practical and physical problems 
regarding dielectric heterogeneous systems like the
radio-frequency and microwave dielectric spectra of living cells \cite{Vrinceanu1996} and 
plasmonic properties of metallic nanoparticles \cite{Ouyang1989,Mayergoyz2005}. 
Another problem is the equilibrium charge 
distribution on a conductor (the Robin problem) \cite{Robin1886} and the implicit 
capacitance with applications in computational biophysics \cite{Simonson2003}, in scanning 
probe microscopy \cite{Hofer2003,Mottaghizadeh2013}, or in electrical charge storage in supercapacitors 
\cite{Simon2008}. 

The capacitance of an arbitrarily shaped body is calculated by ``mimicking'' some directly related 
phenomena like the diffusion-controlled reactions \cite{Douglas1994} or the 
ergodic generation of the equilibrium charge 
distribution \cite{Mascagnia2004}. 
The standard procedures for solving the Laplace equation are the Finite Element 
Method (FEM) \cite{Johnson1987} or the Boundary Integral Equation (BIE) method with the 
finite element formulation as the Boundary Element Method (BEM) 
\cite{Poljak2005}. 
In contrast to the FEM, in the BEM only the surfaces of the inclusions are discretized, such that with 
numerical algorithms 
like the fast multipole method (FMM) of Rokhlin and Greengard \cite{Rokhlin1985,Greengard1987} 
the calculations are essentially of $O(N)$, where $N$ is the number of 
nodes.  
The capacitance of an arbitrary object has been treated in different contexts 
with $O(N)$ FMM schemes (see Ref. \onlinecite{Tausch1998} and the references therein). 
It can be treated as a first- \cite{Nabors1994} or as a second-kind integral equation. 
The second-kind integral formulation is based either on the Neumann-Poincar\'{e} 
operator \cite{Greenbaum1993} or on its adjoint\cite{Tausch1998}. 
The most convenient approach is, however, the second-kind integral equation 
with the adjoint of the Neumann-Poincar\'{e} operator which provides both the charge density and the capacitance \cite{Tausch1998}. 

In this paper we adopt such a BIE method to 
calculate the capacitance and the equilibrium charge on an arbitrarily shaped object with several 
applications. 
The capacitance is obtained concurrently with other physical properties like the static 
and the dynamic polarizabilities of nanoparticles with applications in nanoparticle manipulation \cite{Farajian2008} 
and plasmonics \cite{Mayergoyz2005}.
We use a spectral method \cite{Sandu2010,Sandu2011,Sandu2013} which provides an exponential 
convergence \cite{Boyd2001}. Moreover, our basis functions include spherical harmonics \cite{Sandu2010} that 
can be directly related with the multipoles in the FMM  of Rokhlin \cite{Rokhlin1985,Greengard1987}.      
Compared to others the present method shows directly that the geometric dependence of 
capacitance is incorporated in a norm of a given eigenvector of the Neumann-Poincar\'{e} operator. 
This eigenvector norm varies slowly with the geometry hence two scale-invariant defined shape factors can be readily used in the estimation of 
capacitance for arbitrary shapes. Furthermore, we define a specific class of
capacitors in which the electrodes are the equipotential surfaces generated by the equilibrium charge on an arbitrarily shaped metallic object 
with applications regarding some analytical results like confocal spheroidal capacitors and finite cylinders. We also provide a second order compact 
capacitance formula for thin and constant-thickness capacitors with other applications referring to membrane capacitance of living cells and charge 
storage in supercapacitors.

The paper is organized as follows. In the second section we define the capacitance in the second-kind integral form. 
Then, we define a general capacitor and a specific class of capacitors in the following section. 
Section 4 describes the numerical method and the applications just mentioned above. A summary is given in the last section.

\section{Capacitance of a metallic object in a second-kind integral formulation}

We assume an arbitrarily shaped domain $\Omega $ bounded by 
the surface $\Sigma $ in the 3-dimensional 
space. The following operators can be defined on $\Sigma $: $\hat {M}$, its adjoint $\hat {M}^\dag $, 
and $\hat {S}$.\cite{Kellog1967,Khavinson2007} The action of  $\hat {M}$ on a function $u$ signifies the 
normal electric field to $\Sigma$ 
generated by the charge density $u$. 
The operator $\hat {M}^\dag$, which is the Neumann-Poincar\'{e} (or double-layer) operator, acts on the dipole density $v$ generating an electric potential on $\Sigma$. On the other hand, $\hat {S}$ is a Coulomb (single-layer) operator which acts on the charge density $u$ creating an electric potential on $\Sigma$. 

The operators $\hat {M}$ and $\hat {M}^\dag $ have the same spectrum within $-1/2$ and $1/2$ and the eigenfunctions $u_i $ of $\hat {M}$ are 
related to the eigenfunctions $v_i $ of $\hat {M}^\dag $ by $v_i = \hat 
{S}\left[ {u_i } \right]$, which makes them bi-orthogonal, i.e.,  
$\langle {v_j }|{u_i }\rangle = \delta _{ij} $ \cite{Sandu2013}. 
The 
largest eigenvalue of $\hat {M}$ and $\hat {M}^\dag $ is 1/2 irrespective of 
the domain shape \cite{Khavinson2007,Sandu2010} and the corresponding eigenfunction $v_1 $ of $\hat 
{M}^\dag $ is a constant function, i. e., $v_1 = $constant on $\Sigma $.
As we will discuss below, the companion eigenfunction $u_1 $ of $\hat {M}$ is 
proportional to the equilibrium charge distribution on a conductor of shape 
determined by $\Sigma $. We note 
that the spectrum of $\hat {M}$ and $\hat {M}^\dag $ is scale invariant, but the 
spectrum of $\hat {S}$ is proportional to the linear size of $\Omega $. 

$\hat {M}$ and $\hat {S}$ can be used in the resolution of many physical problems like the static \cite{Farajian2008} or 
the dynamic object polarizability represented by the  
dielectric spectra of living cells \cite{Vrinceanu1996, Sandu2010} or the optical properties of metallic nanoparticles 
\cite{Ouyang1989,Mayergoyz2005, Sandu2011,Sandu2013}. 
Another closely related issue is the Robin problem of finding the equilibrium charge distribution $u_R$ on a 
conductor of arbitrary shape \cite{Robin1886}. It can be 
cast into an integral equation of the second-kind that has the operator form  
\begin{equation}
\label{eq1}
\hat {M}\left[{u_R } \right] = \frac{1}{2}u_R \left( {\bf{x}} \right),
\end{equation}
with the constraint $\int\limits_{{\bf{x}} \in \Sigma } {u_R ( {\bf{x}} 
)d\Sigma  ( {\bf{x}} ) } = 1$. The constraint can be put in the following form 
$ \langle 1 | {u_R }  \rangle = 1 $,
where $1$ is the constant function of value 1 on $\Sigma $. 
Equation (\ref{eq1}) has the obvious solution $u_R \propto u_1 $. The constant value $V_R $ of the electric 
potential generated by $u_R $ is formally given by
$\hat {S}\left[ {u_R } \right] = V_R 1$
and is called the Robin constant, while its inverse is the capacitance C of 
the body bounded by $\Sigma $. If we consider the dielectric permittivity $\varepsilon $ of 
the embedding medium and the constraint $\langle 1| {u_R } \rangle = 1 $ the capacitance is 
$C ={\varepsilon }/{\langle {u_R | {\hat {S}[ {u_R } ]} \rangle } }$.
Furthermore, if $u_R = a_1 u_1$, one can prove that the constant $a_1 $ is 
the proportionality factor between $v_1 $ and 
the constant distribution $1$, i. e., $1 = {v_1 } / {a_1 }$. Then we can relate $a_1 $ to the norm of $v_1 $ by the following 
chain of equations $\left\langle {v_1 \left| {v_1 } \right\rangle } \right. = \left\| {v_1 } 
\right\|^2 = a_1^2 \left\langle {1\left| 1 \right\rangle } \right. = a_1^2 A$,
where $A$ is the area of $\Sigma $. Finally, the capacitance takes a simple and compact form 
\begin{equation}
\label{eq2}
C = \frac{\varepsilon A}{\left\| {v_1 } \right\|^2}.
\end{equation}
One can show that $\left\| {v_1 } \right\|^2$ is proportional to the linear size of 
$\Omega $ therefore, the capacitance itself is also proportional to the linear size of 
the body. Equation (\ref{eq2}) shows explicitly both the geometric dependence of 
capacitance of an arbitrarily shaped object and the scale invariance of the shape factor $C / \sqrt{4 \pi A }$. 
The shape factor varies slowly with the conductor shape \cite{Chow1982} 
hence, as we will discuss in the next section, $\left\| {v_1 } \right\|^2$ is a slowly varying 
function of the conductor shape and perturbative estimations of capacitance can be performed. 
\begin{figure}
 \begin{center}
\includegraphics [width=2 in] {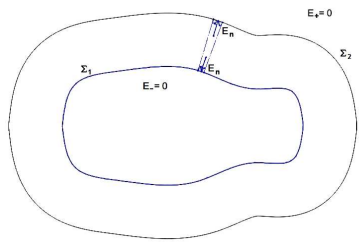}
  \end{center}
\caption{\label{fig:1}
Schematic representation of a capacitor with the configurations of fields inside $\Sigma _1 $ and 
outside  $\Sigma _2 $. The dotted lines delimitate a Gaussian surface used in the text.}
\end{figure}

\section{Capacitors and their capacitance}

\subsection{Definition}
In general, a capacitor 
consists of two separated conducting bodies. We consider a 
capacitor that is made of two smooth surfaces $\Sigma _1 $ and $\Sigma _2 
$ in which $\Sigma _2 $ encloses $\Sigma _1 $ (Fig. \ref{fig:1}). The capacitance 
of the capacitor is defined as the total charge that is held on $\Sigma _1 $ when 
the electrical potential is 1 on $\Sigma _1 $ and 0 on $\Sigma _2 $. Therefore, 
the electrostatic problem is the Laplace equation in the 
space $\Omega _{12} $ between $\Sigma _1 $ and $\Sigma _2 $:
\begin{equation}
\label{eq26}
 \Delta u( {\bf{x}}) = 0;\;{\bf{x}} \in \Omega _{12} 
\end{equation}
with the boundary conditions $u\left( {\bf{x}} \right) = 1$ for ${\bf{x}} \in \Sigma _1$ and 
$u\left( {\bf{x}} \right) = 0$ for ${\bf{x}} \in \Sigma _2$. It is easy to see that with these boundary conditions 
the solution of the Laplace equation 
inside of $\Sigma _1 $ and outside of $\Sigma _2 $ is the constant 1 and the 
constant 0, respectively. Inside $\Omega _{12} $ we seek a 
solution for (\ref{eq26}) in the form of two single-layer 
potentials 
\begin{equation}
\label{eq27}
u\left( x \right) = \int\limits_{
 {\bf{y}} \in \Sigma _1  } {\frac{\mu _1 \left( {\bf{y}} \right)}{4\pi \left| {\bf{x}} - {\bf{y}} \right|}d\Sigma 
\left( {\bf{y}} \right)} + \int\limits_{
 {\bf{y}} \in \Sigma _2 } {\frac{\mu _2 \left( {\bf{y}} \right)}{4\pi \left| {\bf{x}} - {\bf{y}} \right|}d\Sigma 
\left( {\bf{y}} \right)} ,
\end{equation}
where $\mu _1 $ and $\mu _2 $ are the induced charge densities on $\Sigma _1 
$ and $\Sigma _2 $. Similar to $\hat {M}$ and $\hat {S}$ we define on $\Sigma _1 $ and $\Sigma _2$ 
four operators $\hat {M}_{ij} $ and four operators  $\hat {S}_{ij} $ as follows
\begin{equation}
\label{eq28}
\hat {M}_{ij} \left[ {\mu _j } \right] = \int\limits_{
 {\bf{x}} \in \Sigma _i  
 {\bf{y}} \in \Sigma _j  
 } {\frac{\mu _j \left( \bf{y} \right) \bf{n}\left( \bf{x} \right) \cdot \left( 
{\bf{x} - \bf{y}} \right)}{4\pi \left| {\bf{x} - \bf{y}} \right|^3}d\Sigma \left( \bf{y} \right)}, 
\end{equation}

\begin{equation}
\label{eq31}
\hat {S}_{ij} \left[ u \right] = \int\limits_{
 {\bf{x}} \in \Sigma _i   
 {\bf{y}} \in \Sigma _j  
 } {\frac{u\left( {\bf{y}} \right)}{4\pi \left| {\bf{x}} - {\bf{y}} \right|}d\Sigma 
\left( {\bf{y}} \right)},
\end{equation}
with $i,j = \overline {1,2} $. In Eq. (\ref{eq28}) \textbf{n} is the normal vector to $\Sigma_{1,2} $. The equations obeyed by $\mu _1 $ and $\mu _2 
$ are 
\begin{equation}
\label{eq29}
\begin{array}{l}
 \hat {M}_{11} \left[ {\mu _1 } \right] + \hat {M}_{12} \left[ {\mu _2 } 
\right] = \frac{1}{2}\mu _1 \\ 
 \hat {M}_{21} \left[ {\mu _1 } \right] + \hat {M}_{22} \left[ {\mu _2 } 
\right] = - \frac{1}{2}\mu _2 \\ 
 \hat {S}_{11} \left[ {\mu _1 } \right] + \hat {S}_{12} \left[ {\mu _2 } 
\right] = 1 \\ 
 \hat {S}_{21} \left[ {\mu _1 } \right] + \hat {S}_{22} \left[ {\mu _2 } 
\right] = 0. \\ 
 \end{array}
\end{equation}

\noindent
The first two equations of (\ref{eq29}) set the normal fields on $\Sigma _1 $ from inside and on $\Sigma _2 $ from 
outside to zero, while the last two equations are the boundary conditions of (\ref{eq26}). The solution of the first two equations in (\ref{eq29}) is 
the solution of (\ref{eq26}) up to multiplicative constants. The multiplicative constants are fixed 
by the last two equations of (\ref{eq29}).

\subsection{A special class of capacitors}
The capacitance of the capacitor is the total charge on $\Sigma_1$ and 
depends on 
inter-surface operators $\hat {M}_{ij}$ and $\hat {S}_{ij}$. 
In the special case when $\Sigma _2 $ 
is an equipotential surface determined by the equilibrium charge distributed 
on $\Sigma _1 $ a compact capacitance formula can be deduced with the help of $\hat {M}$ and $\hat {S}$ only. 
It is not hard to see that solutions $\mu _1 $ and $\mu _2 $ of the first two equations of (\ref{eq29}) 
are proportional to the equilibrium 
charge densities on $\Sigma _1 $ and $\Sigma _2 $, respectively. To determine $\mu _1 $ and $\mu _2 $ one needs the boundary 
conditions given by the last two equations of (\ref{eq29}). 
Thus, by integrating the third 
equation of (\ref{eq29}) on $\Sigma _1 $ and the fourth equation on $\Sigma _2 $ one obtains the following relations 
$V_1 + V_2 = 1$ and $V_{12} + V_2 = 0$, where $V_1 $ is the electric potential 
induced by $\mu _1 $ on $\Sigma _1 $, $V_2 $ is the electric potential 
induced by $\mu _2 $ inside $\Sigma _2 $ as well as on $\Sigma 
_1 $, and $V_{12} $ is the electric 
potential induced on $\Sigma _2 $ by $\mu _1 $. On the other hand, the total 
charges on $\Sigma _1 $ and on $\Sigma _2 $ are $Q_1  = C_1 V_1$ and $Q_2  =  C_2 V_2$, which 
are valid only if $\Sigma _2 $ 
is one of the equipotential surfaces determined by an equilibrium charge distributed 
on $\Sigma _1 $. Equation (\ref{eq2}) provides the expressions of $C_1 $ and $C_2 $ 
that are the capacitances of $\Sigma _1 $ and $\Sigma _2 $, respectively. 
Keeping in mind that $Q_1 + Q_2 = 0$ we 
obtain the capacitance
\begin{equation}
\label{eq37}
C_{cond} =  \left( {\frac{1}{C_1 } - \frac{1}{C_2 }} \right)^{ - 1}.
\end{equation}

In the limiting case of very thin capacitors (i. e., $\Sigma_2$ being very close to $\Sigma_1$) Eq. (\ref{eq37}) takes 
a planar-like capacitor expression given by 
\begin{equation}
\label{eqthincapacitor}
C_{thin\_capacitor} = \varepsilon \int\limits_{y \in \Sigma } 
{\frac{d\Sigma \left( y \right)}{\delta d}},
 \end{equation}   
where $\delta d$ is the "distance" between  $\Sigma_1$ and $\Sigma_2$ locally defined below. In the vicinity 
of $\Sigma_1$ a coordinate system $(\zeta_1,\zeta_2,\zeta_3)$ can be defined, such that $\zeta_1$ and $\zeta_2$ 
describe $\Sigma_1$ while $\zeta_3$ is the electric potential following the field lines 
from $\Sigma_1$ to $\Sigma_2$. The electric potential $V_2$ on $\Sigma_2$ is related 
to $V_1$, the electric potential on $\Sigma_1$, by 
\begin{equation}
\label{eqdeltapotential}
V_2 \cong V_1 + \frac{\partial V}{\partial \zeta_3} \delta \zeta_3,
\end{equation}
where $\delta \zeta_3$ is a small variation of $\zeta_3$ from  $\Sigma_1$ to $\Sigma_2$. The local thickness 
of the capacitor is $\delta d = h_{\zeta_3} \delta \zeta_3$, with $ h_{\zeta_3}$ as the Lam\'{e} 
coefficient corresponding to $\zeta_3$ \cite{Landau1984}. 
From Eq. (\ref{eqdeltapotential}) Gauss theorem (see for instance Fig. \ref{fig:1}) provides the charge density 
$\sigma = {\varepsilon (V_1 - V_2)}/{\delta d}$. Integrating the charge 
$\sigma$ over $\Sigma_1 $ and dividing by $V_1 - V_2$ one obtains (\ref{eqthincapacitor}). 
Now we consider without loss of generality  that $(\zeta_1,\zeta_2,\zeta_3)$ is orthogonal. 
Then using Eq.  (\ref{eqthincapacitor}) the form of Eq. (\ref{eq37}) can be recast as 
\begin{equation}
\label{eqcapacitor}
C_{cond} = \varepsilon (\int\limits_{1}^{0} \frac{d \zeta_3}{\int\limits_{y \in \Sigma } 
{\frac{ d\Sigma \left( y \right) }{h_{\zeta_3}(y)}}})^{-1}.
 \end{equation} 
The validity 
of (\ref{eqthincapacitor}) is more general than that of the case considered above (in which $\Sigma_2$ is 
an equipotential surface generated by the equilibrium charge on $\Sigma_1$). Some examples will be provided in the next section, where it will be 
also discussed cases in which (\ref{eqthincapacitor}) may not be good enough.    

Particular examples of Eqs. (\ref{eq37}) and (\ref{eqcapacitor}) are the capacitances of 
concentric spheres and of coaxial cables. 
For a capacitor made of two concentric spheres the capacitance is 
$C_{sph\_cond}={4\pi \varepsilon R_1R_2}/{(R_2-R_1)}$, where $R_1$ and $R_2$ are the 
radii of the two spheres with $R_2 > R_1$. Since the capacitance of a sphere 
is $C_{sph}= 4 \pi \varepsilon R$, 
it is easy to check 
that $C_{sph\_cond}$ has the form given by Eq. (\ref{eq37}). 

The capacitance of a capacitor 
made of two confocal spheroids obeys also (\ref{eq37}) and can be calculated with Eq. (\ref{eqcapacitor}). 
Two confocal spheroids are conveniently described in spheroidal coordinates $\left( {\eta ,\xi ,\varphi } \right)$, which 
for prolate spheroids obey the equations
\begin{equation}
\label{eqcoord}
\begin{array}{l}
 x = c\sqrt {\eta ^2 - 1} \sqrt {1 - \xi ^2} \cos \left( \varphi \right) \\ 
 y = c\sqrt {\eta ^2 - 1} \sqrt {1 - \xi ^2} \sin \left( \varphi \right) \\ 
 z = c\eta \xi. \\ 
 \end{array}
\end{equation} 
The two confocal spheroids defining $\Sigma _1 $ and $\Sigma _2 $  are determined by $\eta = \eta _1 $ 
and $\eta = \eta _2 $, respectively. The coordinates $\left( {\eta ,\xi ,\varphi } \right)$ are orthogonal and it 
can be shown that the equipotential surfaces of the eqiulibrium charge on the spheroid of equation $\eta = \eta_1$ is 
any confocal spheroid of equation $\eta = \eta_2 > \eta_1$ \cite{Landau1984}. From Eq. (\ref{eqcapacitor}) 
one obtains directly the capacitance of a confocal spheroidal capacitor 
as in the following expression 
\begin{equation}
\label{eqprolcapacitor}
C_{prolate\_capacitor} = \frac{4 \pi \varepsilon c}{Q_0^0 \left( {\eta _1 } \right) - Q_0^0 \left( \eta_2 \right)},
\end{equation}
where $Q_0^0\left( {\eta} \right) = ln(({\eta +1})/({\eta - 1}))$. Eq. (\ref{eqprolcapacitor}) is of form 
(\ref{eq37}) since the capacitance of a prolate spheroid alone $(\eta_2 \to \infty)$ is 
$C_{prolate\_spheroid} = {4 \pi \varepsilon c}/{Q_0^0 \left( \eta_1 \right)}$ 
which is given in the standard textbooks of classical electrodynamics \cite{Landau1984}.  The capacitances of oblate confocal spheroids are found replacing 
$\eta$ by $i\eta$ and $c$ by $-i c$. An expression 
similar to (\ref{eqprolcapacitor}) was found in a recent paper \cite{Momoh2009}, where the authors did not notice the 
significance of Eq. (\ref{eqprolcapacitor}) in terms of Eq. (\ref{eqcapacitor}). 
\begin{figure}[htp]
  \begin{center}
    \includegraphics [width=3in]{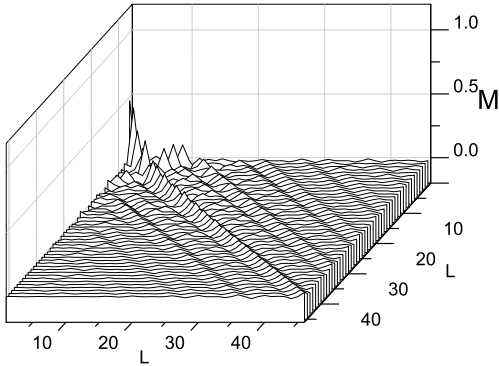}
  \end{center}
\caption{Matrix elements of $\hat{M}$  in the $\tilde{Y}_{Lm}$ basis. The matrix indices designate the indices $L$ of the spherical harmonics $\tilde{Y}_{Lm}$. For axisymmetric objects only $m=0$ is relevant.}
\label{fig:2}
\end{figure}

\section{Applications and Discussion}
\label{}
\subsection{A Numerical method}
The capacitance can be determined in numerical simulators used to calculate other 
physical properties like the dynamic polarizabilities needed for localized plasmon resonances in metallic 
nanoparticles \cite{Ouyang1989,Mayergoyz2005}. Simultaneous calculations of capacitance and polarizability 
were also performed in the path integral formulation by averaging over random walk trajectories \cite{Mansfield2001}.
Our numerical method is an operator based BIE method that 
calculates the eigenvalues $\chi_k$ and the eigenvectors $u_k$ and $v_k$ of $\hat M$ and $\hat M^\dag$, 
respectively. In order to have normed $u_k$ and $v_k$ one needs also to calculate the matrix elements of $\hat S$ \cite{Sandu2013}. 
The present method belongs to the class of 
the spectral methods which are fast converging \cite{Boyd2001}. In these methods the functions of the basis set 
are defined globally rather than locally like in the standard FEM. In our approach the function set is related to the spherical harmonics 
$Y_{Lm} (\theta ,\varphi)$ defined on a sphere that is related to $\Sigma$ by the map 
${\bf x} \to (\theta ({\bf x}),\varphi ({\bf x}))$. Details of the method for axisymmetric 
objects are given in Refs. \onlinecite{Sandu2010,Sandu2011}. In Fig. \ref{fig:2} we plotted the matrix elements of $\hat{M}$ for a generic axially symmetric object. 
It is easy to notice that the matrices 
are sparse with the significant matrix elements being around the diagonal or/and at low-value indices which are basically low-order multipoles. 
The matrix elements of $\hat{S}$ show also a similar behavior. Thus, 
our method is similar to the FMM in which the major contributions come from the low-order multipoles \cite{Rokhlin1985,Greengard1987}.
The validity of the numerical method was checked by calculations performed on oblate and prolate spheroids, which have analytical expressions discussed above. Our numerical calculations show a very good agreement with the analytical results. 
The relative error is at most $5 \times 10^{-5}$ with a relative small overhead of 25 functions in the basis and 
96 quadrature points. The implementation of BIE for axisymmetric shapes has also shown to provide very accurate results of 
the depolarization factors which are related to other eigenvalues of $\hat M$ and $\hat M^\dag$ \cite{Sandu2012}. 

\begin{figure}
  \begin{center}
  \includegraphics [width=3.in]{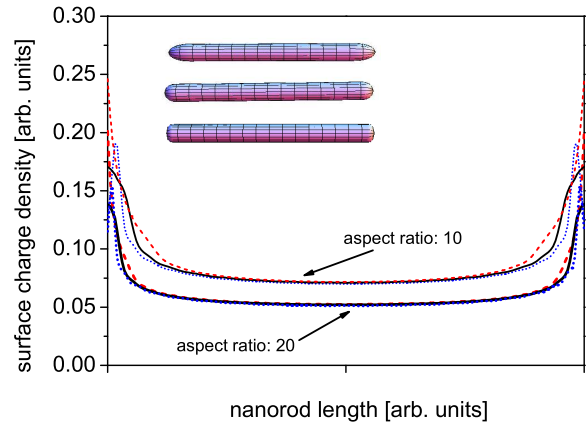}  
  \end{center}
\caption{(Color online) The equilibrium surface charge density on rods with various cappings shown in the inset: hemisphere (black solid line), oblate 
hemispheroid (blue dotted line), and prolate hemispheroid (red dashed line).}
\label{fig:3}
\end{figure}

\subsection{Cylindrical rods}

We have also performed numerical calculations on cylindrical rods with different end-cap geometries: half of an oblate spheroid 
with $1/2$ aspect ratio, half of a sphere, and half of a 
prolate spheroid with an aspect ratio of $2$. In Fig. \ref{fig:3} 
we have plotted the surface charge density of metallic rods  with the above cappings. Two aspect 
ratios have been considered: 10 and 20. 
Since the charge preserves the axial symmetry only the longitudinal dependence is shown. Fig. \ref{fig:3} 
illustrates that, ignoring the capping zones, the charge density is almost the same. The resemblance is greater 
as the aspect ratio becomes larger. In addition, in the middle of the rod the charge density is almost constant and decreases 
with the increase of the aspect ratio.  

Let us now consider two hemispherically capped rods. The first one is determined by the surface $\Sigma_1$ that is a cylinder of length $L$ and 
caps of radius $R_1$. The other rod is determined by the outer surface $\Sigma_2$ with the same 
length $L$, but with a radius $R_2 > R_1$. 
$\Sigma_1$ and $\Sigma_2$ have the same normal hence, 
geometrical intuition tells us that we can apply Eq. (\ref{eqcapacitor}) to obtain the capacitance of a such capacitor made of two finite cylindrical rods 
with hemispherical ends. Explicit numerical calculations of equipotential surfaces
show that the above assumption is quite good. Therefore, 
by applying Eq. (\ref{eqcapacitor}) we obtain the following expression 
\begin{equation}
\label{eqcaprod}
C_{rod\_capacitor} = \frac{2\pi \varepsilon L}{ln(\frac{R_2(R_1+L/2)}{R_1(R_2+L/2)})}.
 \end{equation} 
Eq. (\ref{eqcaprod}) recovers known results in two limiting cases: (a) 
concentric spheres, $L \to 0$; and (b) coaxial cable, $L \to \infty$. In the limiting case of $R_2 \to  \infty$ one obtains the capacitance of 
a finite cylindrical rod with hemispherical cappings 
\begin{equation}
\label{eqrod}
C_{hemispherical\_rod} =\frac{4\pi \varepsilon R_1 (m-1)}{ln(m)},
 \end{equation} 
with $m = ({L+2R_1})/({2R_1})$ as the aspect ratio of the rod. 

This result can be extended to finite cylinders with 
hemispheroidal cappings where the equipotential surfaces are determined by the corresponding confocal spheroids. 
In the particular case of oblate hemispheroidal cappings one may obtain the limit of cylinder with flat cappings. Thus, 
after some tedious but otherwise straightforward calculations the capacity 
of cylinders with flat endings is 
\begin{equation}
\label{eqflatrod}
C_{flat\_rod} =\frac{4\pi \varepsilon R (m^2+1)}{m(ln(m) + \frac{\pi}{2m})},
 \end{equation} 
with $m = {L}/({2R})$ as the aspect ratio of the rod. Here $L$ is the length of the rod and $R$ is its axial radius. We easily notice that 
for $m \to 0$  Eq. (\ref{eqflatrod}) reproduces the capacitance of a disk \cite{Landau1984}.
\begin{figure}[htp]
  \begin{center}
   \subfigure {\label{fig:4a} \includegraphics [width=3.in]{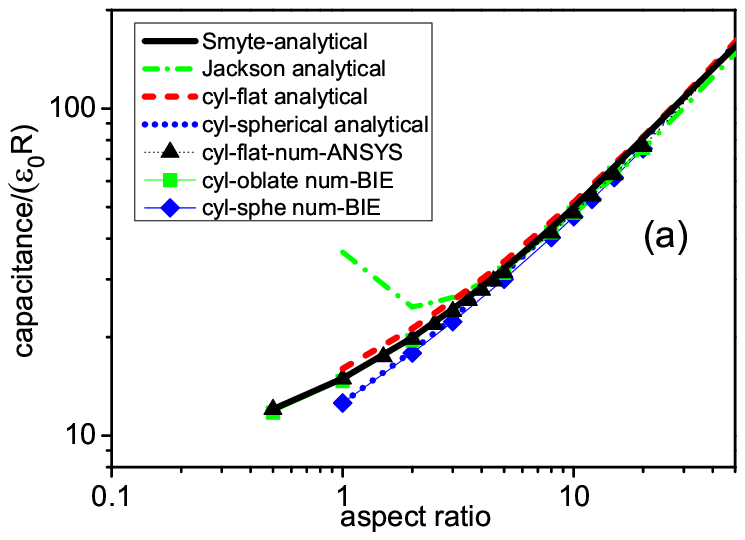}}  
    \subfigure {\label{fig:4b} \includegraphics [width=3.in]{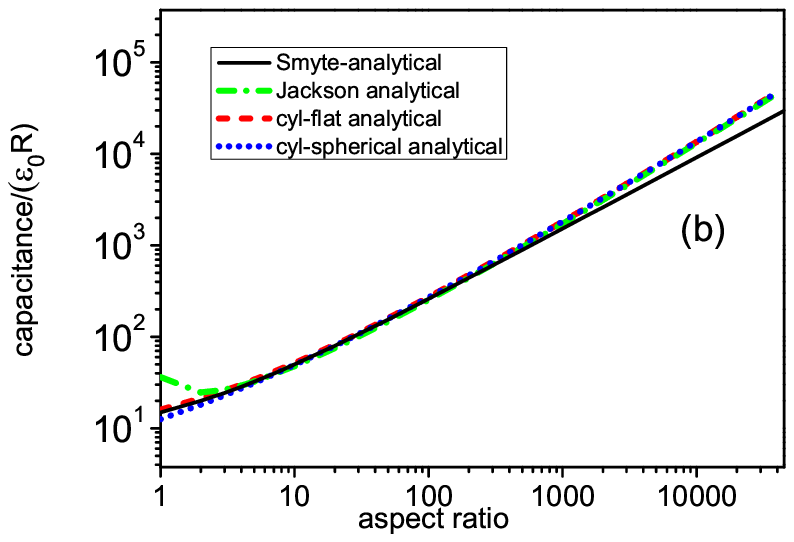}} 
  \end{center}
\caption{(Color online) (a) Analytical and numerical calculations of capacitance for various cylindrical rods 
at relatively small aspect ratios. (b) Various analytical results over a sufficiently large range of aspect ratios. 
The capacitances are given in terms of $\varepsilon_0R$, with $\varepsilon_0$ as the vacuum permittivity 
and $R$ as the cross-sectional radius.}
\label{fig:4}
\end{figure}

 We have compared numerical capacitance calculations of cylindrical rods having different capping geometries (flat, oblate hemispheroidal, 
and hemispherical) with Eqs. (\ref{eqrod}) and (\ref{eqflatrod}) and with Smythe's \cite{Smythe1962} and Jackson's \cite{Jackson2000} analytical results. 
Numerical results are obtained using either the BIE method for rods with oblate hemispheroidal and hemispherical cappings or 
the multi-physics program ANSYS (found at wwww.ansys.com) for rods with flat cappings. The results are given in Fig. \ref{fig:4}a. 
The oblate hemispheroidal cappings are chosen to be thin (an aspect ratio of 10), which provides quite 
good approximants for cylinders with flat ends. For example, the capacitances of flat and oblate hemispheroidal capped cylinders are 
apart only by $2\%$ at an aspect ratio of $1/2$.  Moreover, for aspect ratios 
greater than 5 the capacitance of the rods do not depend any longer on the end-cap geometry (the differences are well below $1\%$). On the other hand, the 
analytical results of Eq. (\ref{eqrod}) are apart by up to 5.5\% from the BIE calculations for cylinders with hemispherical 
ends at the aspect ratio of 20. Furthermore, Eq. (\ref{eqflatrod}) is also within a few percentage points from the exact results at 
low aspect ratios, 
but at  larger aspect ratios Eqs. (\ref{eqrod}) and (\ref{eqflatrod}) are sufficiently close. In Fig. \ref{fig:4}b we compare Eqs. (\ref{eqrod}) 
and (\ref{eqflatrod}) with the Smythe's \cite{Smythe1962} and Jackson's \cite{Jackson2000} formulae.  There are known that the Smythe's formula  
is valid at 
low aspect ratios (below 10) \cite{Smythe1962}, while the Jackson's \cite{Jackson2000} works well at large aspect ratios. 
We mention that in another derivation there were obtained also two different expressions of capacitance for short and the long cylinders, 
respectively \cite{Butler1980}.
In contrast, as one can see from Fig. \ref{fig:4} Eqs. (\ref{eqrod}) and (\ref{eqflatrod}) work well for both short and the long cylinders and 
 also have simple algebraic expressions. These results are not that surprising after all. They are asymptotically exact for $m \to 0$ and $m \to \infty$
 by construction. The case of very long cylinders was first considered by Maxwell \cite{Maxwell1877}, who stated that, for $m \to \infty$, the 
charge density tends to be constant. As proved in a separate paper \cite{Jackson2002} the charge and implicitly the capacitance given 
by Jackson \cite{Jackson2000} are similar to those of Maxwell's (e.g., $C=4 \pi \varepsilon Rm/(ln(4m)-1)$) which are accurate for large but not for small $m's$. 

\begin{figure}[htp]
 \begin{center}
 \includegraphics [width=3 in] {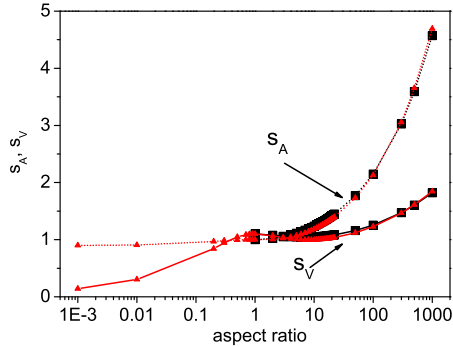}
 \end{center}
   \caption{\label{fig:5}
(Color on-line) Scale-invariant shape factors versus aspect ratio. The shape factors are defined either by the area of the 
object ($s_A$-dotted lines with symbols) or by its volume ($s_V$-solid lines with symbols). 
The spheroids and the cylidrical rods are denoted by black squares and by red triangles, respectively.}
\end{figure}

\subsection{Scale-invariant shape factors and some consequences}

Since the capacitance is proportional to the linear size of the object one can define shape factors that are scale-invariant. For 
instance one can employ the surface or the volume of the object to define scale-invariant shape factors. Two of such shape factors are presented 
in Fig. \ref{fig:5} for cylinders and spheroids. 

The first scale-invariant shape 
factor defined as $s_A = A^{1/2}/(2 \pi^{1/2}||v_1||^2)$ is related to the area $A$ of the object, such that it becomes 1 for spherical shape. 
It shows a relative shape insensitivity for aspect ratios less than 5 and for flat structures. This shape factor has 
been used in isoperimetric inequalities to estimates the capacitance of objects with shapes close to the spherical shape \cite{Chow1982}. 

The second scale-invariant shape factor related to capacitance is defined by $s_V = V^{1/3}/(\pi^{1/3}||v_1||^2)$. It is determined by the volume $V$ of the 
object and shows 
shape insensitivity for long structures (Fig. \ref{fig:5}). 
Thus for aspect ratios from 5 to 40  $s_V$ is almost 1 for both rods 
and spheroids. In contrast  to $s_A$,  $s_V$ varies not much for long structure, but it goes to 0 for flat structures.

These shape factors can be straightforwardly utilized in approximate capacitance calculations for metallic object of various shapes. For instance, 
the capacitance for smooth shape approximants of the object can be calculated with the BIE method and then we may amend the final result with the appropriate area or volume by considering that the shape factor remains unchanged.  

Our findings explain the $V^{1/3}$ scaling found for quantum capacitance of molecular nanowires \cite{Ellenbogen2007},
at least for the aspect ratios $m$ from $5$ to $30$. For long rods, on the other hand, the volume $V$ and the area $A$ of the rod scale almost linearly with $m$. But the quantum capacitance is in fact proportional to $ln(m)$ (see Ref. \onlinecite{Ellenbogen2007}), 
which turns out to be quite close to $m^{1/3}$ for $m$ between 5 and 30. This simple remark explains the $V^{1/3}$ scaling 
of quantum capacitance in the same range of aspect ratios. So, for long structures 
the volume plays a greater role in determining both classical and quantum capacitances.   
\begin{figure}
  \begin{center}
    \subfigure {\label{fig:6a} \includegraphics [width=3.in]{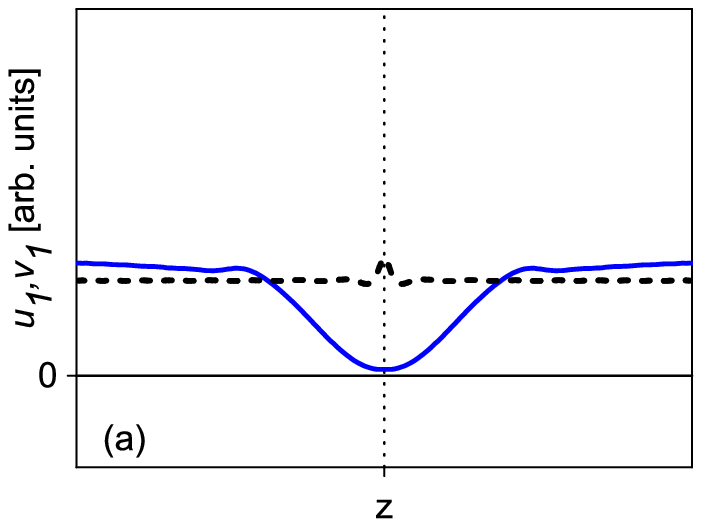}} 
    \subfigure {\label{fig:6b} \includegraphics [width=3.in]{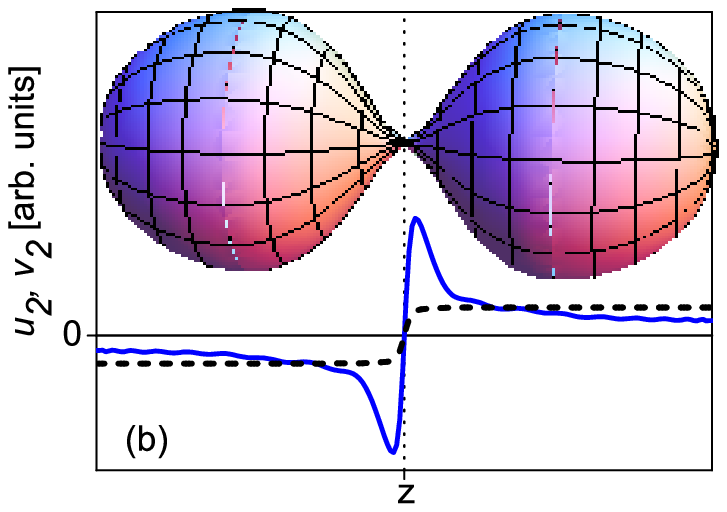}}  
  \end{center}
\caption{(Color online) (a) The axial (z-) dependence of the first and (b) of the second eigefunctions of $\hat{M}$ and $\hat{M}^\dagger$ for a dimer; $u_1$ and $u_2$ are plotted with blue solid 
line, while $v_1$ and $v_2$ are depicted with black dashed lines. The inset of (b) shows the actual dimer. }
\label{fig:6}
\end{figure}

An interesting case is that of a dimer with slightly connected metallic particles of nearly spherical shape \cite{Sandu2011,Sandu2013} with the shape 
depicted in the inset of Fig.  \ref{fig:6}b. 
In these systems the area and the volume related shape factors should become equivalent or close to that since they scales as $R^2$ and $R^3$, respectively, $R$ being the radius 
of the particles (the particles in the dimer are not quite spheres but very close that). As a result the capacitance of the dimers scales with the 
radius of the constituent particles. Our numerical calculations show that both the area and the volume 
related shape factors are very close to 1, i. e., $s_A=1.009$ and $s_V=0.986$.  Now we consider the case of two just touching spheres 
like that treated in Refs. \onlinecite{Lekner2011,Lekner2012}, where analytical expressions of capacitance are provided. The capacitance of touching spheres is \cite{Lekner2011,Lekner2012}: $C_{dimer}= (2 ln 2) \times 4 \pi \varepsilon R \approx 1.386 \times 4 \pi \varepsilon R$. 
If we consider that $s_A=1$ we obtain a capacitance  $C_{dimerA}= {2}^{1/2} \times 4 \pi \varepsilon R \approx 1.4142 \times 4 \pi \varepsilon R$. 
Similarly if  $s_V=1$ the capacitance is $C_{dimerV}= {(8/3)}^{1/3} \times 4 \pi \varepsilon R \approx 1.387 \times 4 \pi \varepsilon R$. It is easy to 
check 
that the two scale-invariant shape factors also reproduce with a good accuracy the results of the asymmetric dimers given in Ref. \onlinecite{Lekner2011}. 

In dimers made of slightly connected particles many eigenfunctions of $\hat{M}$  and 
$\hat{M}^\dagger$ are hybrid eigenfunctions of the constituent particles \cite{Sandu2011,Sandu2013}. An example is provided in 
Figs. \ref{fig:6}a and \ref{fig:6}b, where the first two eigenfunctions of $\hat{M}$  and $\hat{M}^\dagger$ are plotted. The first eigenfunctions $u_1$ and $v_1$ of $\hat{M}$  and $\hat{M}^\dagger$, respectively are basically symmetric 
combinations of the first eigenfunctions in the constituent particles. At the same time, the second eigenfunctions $u_2$ and $v_2$ are antisymmetric 
combinations of the same first eigenfunctions of the constituent particles.
We notice that the first eigenmode $u_1$ provides the charging while it can not be a plasmon 
mode in metallic nanoparticles \cite{Sandu2013}. In the space between the 
particles of the dimer the charge is repelled thus, the particles themselves repell each other (Fig. \ref{fig:6}a). The second eigenmode, 
however, is a plasmon active mode in the long wavelength range \cite{Sandu2011}. Since the shape of $v_2$ is also constant on each particle, 
the first and the second eigenmode 
of $\hat{M}$ lead to the reminiscence of $C(V,V)$ and $C(Q,-Q)$, respectively, when the particles of the dimer are separated \cite{Lekner2011,Lekner2012b}. 
$C(V,V)$ is the capacitance when the two particles are kept at the same potential $V$ and, at touching, turns into the dimer 
capacitance discused above. In contrast, $C(Q,-Q)$ is the capacitance when the two particles are charged with opposite charges $Q$ and $-Q$. It 
logarithmically diverges as the spheres approach the touching point \cite{Lekner2011,Lekner2012b} since the charging mode transforms into a 
dipole-active mode. This behavior of $C(V,V)$ and $C(Q,-Q)$ is expected to hold for dimers of any shape as it has been recently found for ellipsoidal \cite{Murovec2013} or other, more general \cite{Khair2013}, shaped dimers. In addition, Fig. \ref{fig:6}a is an illustative image of sphere repulsion when they are in 
contact \cite{Lekner2012} and Fig. \ref{fig:6}b provides a glimpse of sphere attraction when they are 
kept at a constant voltage difference \cite{Lekner2012,Lekner2012b}.     

\begin{figure}
  \begin{center}
  \includegraphics [width=3.in]{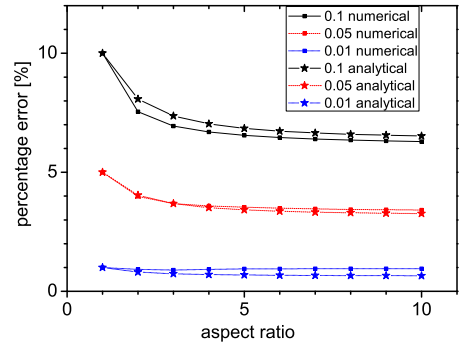}  
  \end{center}
\caption{(Color online) The percentage error of the planar-like capacitor formula with respect to the full numerical calculations (the square symbols) and 
with respect to analytical formula (the star symbols), i. e., Eq. (\ref{eqthincapacitor2}), for thin constant-thickness spheroidal capacitors.}
\label{fig:7}
\end{figure}

\subsection{Thin capacitors}

Another application is the estimation of capacitance in thin capacitors like the membrane capacitance in living cells. The shelled ellipsoidal model and 
the spheroidal model, in particular, are two of the most common models of living cells used in modeling the dielectric spectroscopy 
experiments. In the spheroidal shelled model the shell designates the cell 
membrane, which is practically non-conductive and bounded by two confocal spheroids. Both the spherical \cite{Prodan2008} and 
the spheroidal \cite{Biasio2010} models have analytical solutions for $\alpha$-(below the frequency of 10 KHz) and $\beta$- (in the MHz range of the radiofrequency spectrum) relaxations \cite{Stoy1982}. On the other hand, the membrane capacitance 
needs to be estimated as an input 
parameter in the analysis of $\alpha$- and $\beta$-relaxations \cite{Prodan2008}. 
Eq. (\ref{eqprolcapacitor}) gives us also 
an analytical form of the membrane capacitance in the spheroidal model, putting the ellipsoidal and the spherical 
models on the same footing in terms of solvability. 
However, realistic cell models would imply constant-thickness membrane, which is not found in the spheroidal model. In living cells, on the other hand, 
the membrane is very just a few tens of $nm$ thick for a cell size in the $\mu m$ range. Apparently, Eq. (\ref{eqthincapacitor}) is 
quite general for thin capacitors. The control parameter for accuracy is $\zeta_1$ which is weighted by its Lam\'{e} coefficient. 
Thus, it is of great interest to look for the validity of (\ref{eqthincapacitor}) in the case of thin and constant-thickness capacitors.  

We suppose that $\Sigma_1 $ is sufficiently smooth and let us pick an arbitrary point ${\bf{r}}_0$ on $\Sigma_1 $ that is locally 
parametrized by ${{\bf{r}}(\zeta_1,\zeta_2) }$ and consider its two principal directions in which the curvature tensor is diagonal. 
We assume without loss of generality that $\zeta_1$ is the parameterization of the first principal direction and $\zeta_2$ 
the parameterization of the second. We further assume that the unit tangent vectors are 
${\bf t}_{1,2}  = {\frac{\partial {\bf r}}{\partial \zeta_{1,2} }}$, such that the surface element is $d\Sigma_1=d\zeta_1 d\zeta_2$. 
Hence, the unit normal vector on $\Sigma_1$ at ${\bf{r}}_0$ is
${\bf n} = {\bf t}_{1}  \times {\bf t}_{2} $. 
We also denote the curvature along the first principal
direction at ${\bf{r}}_0$ as $\kappa_1$  and the curvature along the second principal direction at ${\bf{r}}_0$ as $\kappa_2$.  
In this situation the constant-thickness capacitor is determined by $\Sigma_1$ and by $\Sigma_2$ that is defined as 
${{\bf r'}(\zeta_1,\zeta_2) = {\bf r}(\zeta_1,\zeta_2) + \delta a {\bf n}(\zeta_1,\zeta_2) } $, where $\delta a$ is the is small and constant. 
Then the tangent vectors on $\Sigma_2$ are ${\bf t}'_{1,2} = {\bf t}_{1,2}(1+\delta a \kappa_{1,2} )$.
Therefore, in the linear approximation with respect to $\delta a $, the surface element on $\Sigma_2$ is 
$d\Sigma_2= d\zeta_1 d\zeta_2(1+ 2 \delta a H)$, with the mean curvature $H= (\kappa_{1}+\kappa_{2})/2$. 
The key approximation that comes under scrutiny is 
Eq. (\ref{eqdeltapotential}), which is valid as long as the electric field on $\Sigma_2$ is the same as the electric field on $\Sigma_1$. According to 
the Gauss theorem the product of electric field and the surface element must be constant. In the linear approximation with respect to 
$\delta a $ the 
field $E_2$ on $\Sigma_2$ is $E_2= E_1(1 - 2 \delta a H)$, where $E_1$ is the field on $\Sigma_1$. Therefore, it is not hard to see 
that the voltage drop $V_1-V_2$ must 
be amended by the factor $(1 - \delta a H)$ and  Eq. (\ref{eqdeltapotential}) must be changed to 
\begin{equation}
\label{eqthincapacitor2}
C_{membrane} = \varepsilon \frac{A}{\delta a}+  \int\limits_{y \in \Sigma_1 } 
{\varepsilon H d\Sigma_1(y)}.
 \end{equation}   
Eq. (\ref{eqthincapacitor2}) is a second order approximation of thin constant-thickness, where the first term on the right-hand side is the leading term. 
We have two straighforward consequences. 
First, Eq. (\ref{eqthincapacitor2}) explains why the capacitance of cell membranes appear to be much larger 
whenever the membrane is folded \cite{Lo1995}. Second, it gives us also the validity criterion of parallel-plane 
capacitor-like formula which is $ {\delta a \int{H d\Sigma_1(y)}}/{A} \ll 1.$
In Fig. \ref{fig:7} we present the percentage error of the planar-like capacitor formula, Eq.(\ref{eqthincapacitor}), with respect to full 
numerical calculations and with respect to 
Eq. (\ref{eqthincapacitor2}) for thin constant-thickness capacitors of prolate spheroidal shape. Three thicknesses are considered: 0.1, 0.05, and 0.01 of 
the largest axial cross-sectional radius of the spheroid. The calculations show that Eq. (\ref{eqthincapacitor2}) is a good approximation of 
constant-thickness capacitors up to significant thichnesses. On the other hand, for extremely thin capacitors (a thickness of 0.01) the error of Eq.(\ref{eqthincapacitor}) is about $1 \%$.  

A useful application of Eq. (\ref{eqthincapacitor2}) is the assessment of the geometry-dependent energy storage in supercapacitors. 
In a recent paper Huang et al. \cite{Huang2010} noticed that the normalized capacitance (capacitance per unit area) of spherical and cylindrical 
double-layer capacitors increase with  decreasing sphere and cylinder diameters. They also noticed that the capacitances of spherical 
capacitors increase faster than that of cylinders and argued that this behavior is related to the principal curvatures of those shapes \cite{Huang2010}.
 Eq. (\ref{eqthincapacitor2}) provides not only the proof for the curvature-related capacitance but also a quantitative evaluation of shape dependent 
capacitance. In addition, the same equation can be utilized for double-layer capacitors of arbitrary shapes.

\section{Summary}

In the second-kind integral equation based on the adjoint of the  Neumann-Poincar\'{e} operator the capacitance as well as the static and 
dynamic polarizabilities can be simultaneously calculated for arbitrarily shaped objects. A compact capacitance formula is obtained and 
is simply stated as follows.  
The capacitance is direct proportional to the dielectric permittivity of the embedding medium and 
to the area of the object, and inverse proportional to the squared norm of the eigenfunction of the Neumann-Poincar\'{e} operator 
with the largest eigenvalue. 
A spectral based numerical implementation of the method is accurate and resembles the fast multipole method. 
Several applications are discussed.
The capacitance formula allows us to define scale-invariant shape factors that varies slowly with shape and can be used in 
approximate calculations of capacitance. We have analyzed two scale-invariant shape factors. One of 
the shape factors employs the 
volume of the object and is 
more suitable for long shapes like rods or wires. The other shape factor, which is defined in terms of the object area, 
is more appropriate for objects with shapes close to a sphere. Both scale-invariant shape factors, however,  provide an accurate capacitance of 
touching metallic dimers.

We have extended the above results to capacitors. More explicitly, we have considered a special class of capacitors defined by the equipotential 
surfaces of the equilibrium charge on an arbitrarly shaped body. In this case the capacitor behaves 
like a series 
capacitor with the total capacitance as being the capacitance of the inner surface in series with the opposite (negative) 
capacitance of the outer surface of 
the capacitor. This result leads to an integral form of capacitance that was used to estimate analytically the capacitance of 
confocal spheroidal capacitors and of finite cylinders. 
Another consequence is a second-order formula for thin constant-thickness capacitors of arbitrary shape. The first order term has a 
plane-capacitor like form, while the second 
order term is the surface integral of the mean curvature.  Applications of a thin constant-thickness capacitor formula are encountered in 
the capacitance estimation 
of membrane in living cells and of supercapacitors with arbitrary shapes. 

\begin{acknowledgments}
This work was supported by a grant of the Romanian National Authority for 
Scientific Research, CNCS -- UEFISCDI, project number PNII-ID-PCCE-2011 
-2-0069. 
\end{acknowledgments}


\begin{thebibliography}{45}
\expandafter\ifx\csname natexlab\endcsname\relax\def\natexlab#1{#1}\fi
\expandafter\ifx\csname bibnamefont\endcsname\relax
  \def\bibnamefont#1{#1}\fi
\expandafter\ifx\csname bibfnamefont\endcsname\relax
  \def\bibfnamefont#1{#1}\fi
\expandafter\ifx\csname citenamefont\endcsname\relax
  \def\citenamefont#1{#1}\fi
\expandafter\ifx\csname url\endcsname\relax
  \def\url#1{\texttt{#1}}\fi
\expandafter\ifx\csname urlprefix\endcsname\relax\def\urlprefix{URL }\fi
\providecommand{\bibinfo}[2]{#2}
\providecommand{\eprint}[2][]{\url{#2}}

\bibitem[{\citenamefont{Kellog}(1967)}]{Kellog1967}
\bibinfo{author}{\bibfnamefont{O.~D.} \bibnamefont{Kellog}},
  \emph{\bibinfo{title}{Foundations of Potential Theory}}
  (\bibinfo{publisher}{Springer-Verlag},
  \bibinfo{address}{Berlin-Heidelberg-New York}, \bibinfo{year}{1967}).

\bibitem[{\citenamefont{Khavinson et~al.}(2007)\citenamefont{Khavinson,
  Putinar, and Shapiro}}]{Khavinson2007}
\bibinfo{author}{\bibfnamefont{D.}~\bibnamefont{Khavinson}},
  \bibinfo{author}{\bibfnamefont{M.}~\bibnamefont{Putinar}}, \bibnamefont{and}
  \bibinfo{author}{\bibfnamefont{H.~S.} \bibnamefont{Shapiro}},
  \bibinfo{journal}{Arch. Ration. Mech. Anal.} \textbf{\bibinfo{volume}{185}},
  \bibinfo{pages}{143} (\bibinfo{year}{2007}).

\bibitem[{\citenamefont{Vrinceanu and Gheorghiu}(1996)}]{Vrinceanu1996}
\bibinfo{author}{\bibfnamefont{D.}~\bibnamefont{Vrinceanu}} \bibnamefont{and}
  \bibinfo{author}{\bibfnamefont{E.}~\bibnamefont{Gheorghiu}},
  \bibinfo{journal}{Bioelectrochem. Bioenerg.} \textbf{\bibinfo{volume}{40}},
  \bibinfo{pages}{167} (\bibinfo{year}{1996}).

\bibitem[{\citenamefont{Ouyang and Isaacson}(1989)}]{Ouyang1989}
\bibinfo{author}{\bibfnamefont{F.}~\bibnamefont{Ouyang}} \bibnamefont{and}
  \bibinfo{author}{\bibfnamefont{M.}~\bibnamefont{Isaacson}},
  \bibinfo{journal}{Philos. Mag. B} \textbf{\bibinfo{volume}{60}},
  \bibinfo{pages}{481} (\bibinfo{year}{1989}).

\bibitem[{\citenamefont{Mayergoyz et~al.}(2005)\citenamefont{Mayergoyz,
  Fredkin, and Zhang}}]{Mayergoyz2005}
\bibinfo{author}{\bibfnamefont{I.~D.} \bibnamefont{Mayergoyz}},
  \bibinfo{author}{\bibfnamefont{D.~R.} \bibnamefont{Fredkin}},
  \bibnamefont{and} \bibinfo{author}{\bibfnamefont{Z.}~\bibnamefont{Zhang}},
  \bibinfo{journal}{Phys. Rev. B} \textbf{\bibinfo{volume}{72}},
  \bibinfo{pages}{155412} (\bibinfo{year}{2005}).

\bibitem[{\citenamefont{Robin}(1886)}]{Robin1886}
\bibinfo{author}{\bibfnamefont{G.}~\bibnamefont{Robin}}, \bibinfo{journal}{Ann.
  Sci. Ecole Norm. Sup.} \textbf{\bibinfo{volume}{3}}, \bibinfo{pages}{1}
  (\bibinfo{year}{1886}).

\bibitem[{\citenamefont{Simonson}(2003)}]{Simonson2003}
\bibinfo{author}{\bibfnamefont{T.}~\bibnamefont{Simonson}},
  \bibinfo{journal}{Rep. Prog. Phys.} \textbf{\bibinfo{volume}{66}},
  \bibinfo{pages}{737} (\bibinfo{year}{2003}).

\bibitem[{\citenamefont{Hofer et~al.}(2003)\citenamefont{Hofer, Foster, and
  Shluger}}]{Hofer2003}
\bibinfo{author}{\bibfnamefont{W.~A.} \bibnamefont{Hofer}},
  \bibinfo{author}{\bibfnamefont{A.~S.} \bibnamefont{Foster}},
  \bibnamefont{and} \bibinfo{author}{\bibfnamefont{A.~L.}
  \bibnamefont{Shluger}}, \bibinfo{journal}{Rev. Mod. Phys.}
  \textbf{\bibinfo{volume}{75}}, \bibinfo{pages}{1287} (\bibinfo{year}{2003}).

\bibitem[{\citenamefont{Mottaghizadeh et~al.}(2013)\citenamefont{Mottaghizadeh,
  Lang, Cui, Lesueur, Li, Zheng, Rebuttini, Pinna, Zimmers, and
  Aubin}}]{Mottaghizadeh2013}
\bibinfo{author}{\bibfnamefont{A.}~\bibnamefont{Mottaghizadeh}},
  \bibinfo{author}{\bibfnamefont{P.~L.} \bibnamefont{Lang}},
  \bibinfo{author}{\bibfnamefont{L.~M.} \bibnamefont{Cui}},
  \bibinfo{author}{\bibfnamefont{J.}~\bibnamefont{Lesueur}},
  \bibinfo{author}{\bibfnamefont{J.}~\bibnamefont{Li}},
  \bibinfo{author}{\bibfnamefont{D.~N.} \bibnamefont{Zheng}},
  \bibinfo{author}{\bibfnamefont{V.}~\bibnamefont{Rebuttini}},
  \bibinfo{author}{\bibfnamefont{N.}~\bibnamefont{Pinna}},
  \bibinfo{author}{\bibfnamefont{A.}~\bibnamefont{Zimmers}}, \bibnamefont{and}
  \bibinfo{author}{\bibfnamefont{H.}~\bibnamefont{Aubin}},
  \bibinfo{journal}{Appl. Phys. Lett.} \textbf{\bibinfo{volume}{102}},
  \bibinfo{pages}{053118} (\bibinfo{year}{2013}).

\bibitem[{\citenamefont{Simon and Gogotsi}(2008)}]{Simon2008}
\bibinfo{author}{\bibfnamefont{P.}~\bibnamefont{Simon}} \bibnamefont{and}
  \bibinfo{author}{\bibfnamefont{Y.}~\bibnamefont{Gogotsi}},
  \bibinfo{journal}{Nature Mat.} \textbf{\bibinfo{volume}{7}},
  \bibinfo{pages}{845} (\bibinfo{year}{2008}).

\bibitem[{\citenamefont{Douglas et~al.}(1994)\citenamefont{Douglas, Zhou, and
  Hubbard}}]{Douglas1994}
\bibinfo{author}{\bibfnamefont{J.~F.} \bibnamefont{Douglas}},
  \bibinfo{author}{\bibfnamefont{H.~X.} \bibnamefont{Zhou}}, \bibnamefont{and}
  \bibinfo{author}{\bibfnamefont{J.~B.} \bibnamefont{Hubbard}},
  \bibinfo{journal}{Phys. Rev. E} \textbf{\bibinfo{volume}{49}},
  \bibinfo{pages}{5319} (\bibinfo{year}{1994}).

\bibitem[{\citenamefont{Mascagnia and Simonov}(2004)}]{Mascagnia2004}
\bibinfo{author}{\bibfnamefont{M.}~\bibnamefont{Mascagnia}} \bibnamefont{and}
  \bibinfo{author}{\bibfnamefont{N.~A.} \bibnamefont{Simonov}},
  \bibinfo{journal}{J. Comput. Phys.} \textbf{\bibinfo{volume}{195}},
  \bibinfo{pages}{465} (\bibinfo{year}{2004}).

\bibitem[{\citenamefont{Johnson}(1987)}]{Johnson1987}
\bibinfo{author}{\bibfnamefont{C.}~\bibnamefont{Johnson}},
  \emph{\bibinfo{title}{Numerical Solutions of Partial Differential Equations
  by Finite Element Method}} (\bibinfo{publisher}{Cambridge Univ. Press},
  \bibinfo{address}{Cambridge}, \bibinfo{year}{1987}).

\bibitem[{\citenamefont{Poljak and Brebbia}(2005)}]{Poljak2005}
\bibinfo{author}{\bibfnamefont{D.}~\bibnamefont{Poljak}} \bibnamefont{and}
  \bibinfo{author}{\bibfnamefont{C.~A.} \bibnamefont{Brebbia}},
  \emph{\bibinfo{title}{Boundary Element Methods for Electrical Engineers}}
  (\bibinfo{publisher}{WIT}, \bibinfo{address}{Boston}, \bibinfo{year}{2005}).

\bibitem[{\citenamefont{Rokhlin}(1985)}]{Rokhlin1985}
\bibinfo{author}{\bibfnamefont{V.}~\bibnamefont{Rokhlin}}, \bibinfo{journal}{J
  Comput. Phys.} \textbf{\bibinfo{volume}{60}}, \bibinfo{pages}{187}
  (\bibinfo{year}{1985}).

\bibitem[{\citenamefont{Greengard and Rokhlin}(1987)}]{Greengard1987}
\bibinfo{author}{\bibfnamefont{L.~F.} \bibnamefont{Greengard}}
  \bibnamefont{and} \bibinfo{author}{\bibfnamefont{V.}~\bibnamefont{Rokhlin}},
  \bibinfo{journal}{J. Comput. Phys.} \textbf{\bibinfo{volume}{73}},
  \bibinfo{pages}{325} (\bibinfo{year}{1987}).

\bibitem[{\citenamefont{Tausch and White}(1998)}]{Tausch1998}
\bibinfo{author}{\bibfnamefont{J.}~\bibnamefont{Tausch}} \bibnamefont{and}
  \bibinfo{author}{\bibfnamefont{J.}~\bibnamefont{White}},
  \bibinfo{journal}{Adv. Comput. Math.} \textbf{\bibinfo{volume}{9}},
  \bibinfo{pages}{217} (\bibinfo{year}{1998}).

\bibitem[{\citenamefont{Nabors et~al.}(1994)\citenamefont{Nabors, Korsmeyer,
  Leighton, and White}}]{Nabors1994}
\bibinfo{author}{\bibfnamefont{K.}~\bibnamefont{Nabors}},
  \bibinfo{author}{\bibfnamefont{F.~T.} \bibnamefont{Korsmeyer}},
  \bibinfo{author}{\bibfnamefont{F.~T.} \bibnamefont{Leighton}},
  \bibnamefont{and} \bibinfo{author}{\bibfnamefont{J.}~\bibnamefont{White}},
  \bibinfo{journal}{SIAM J. Sci. Statist. Comput.}
  \textbf{\bibinfo{volume}{15}}, \bibinfo{pages}{713} (\bibinfo{year}{1994}).

\bibitem[{\citenamefont{Greenbaum et~al.}(1993)\citenamefont{Greenbaum,
  Greengard, and Fadden}}]{Greenbaum1993}
\bibinfo{author}{\bibfnamefont{A.}~\bibnamefont{Greenbaum}},
  \bibinfo{author}{\bibfnamefont{L.}~\bibnamefont{Greengard}},
  \bibnamefont{and} \bibinfo{author}{\bibfnamefont{G.~B.~M.}
  \bibnamefont{Fadden}}, \bibinfo{journal}{J. Comput. Phys.}
  \textbf{\bibinfo{volume}{105}}, \bibinfo{pages}{267} (\bibinfo{year}{1993}).

\bibitem[{\citenamefont{Farajian et~al.}(2008)\citenamefont{Farajian,
  Pupysheva, Schmidt, and Yakobson}}]{Farajian2008}
\bibinfo{author}{\bibfnamefont{A.~A.} \bibnamefont{Farajian}},
  \bibinfo{author}{\bibfnamefont{O.~V.} \bibnamefont{Pupysheva}},
  \bibinfo{author}{\bibfnamefont{H.~K.} \bibnamefont{Schmidt}},
  \bibnamefont{and} \bibinfo{author}{\bibfnamefont{B.~I.}
  \bibnamefont{Yakobson}}, \bibinfo{journal}{Phys. Rev. B}
  \textbf{\bibinfo{volume}{77}}, \bibinfo{pages}{205432}
  (\bibinfo{year}{2008}).

\bibitem[{\citenamefont{Sandu et~al.}(2010)\citenamefont{Sandu, Vrinceanu, and
  Gheorghiu}}]{Sandu2010}
\bibinfo{author}{\bibfnamefont{T.}~\bibnamefont{Sandu}},
  \bibinfo{author}{\bibfnamefont{D.}~\bibnamefont{Vrinceanu}},
  \bibnamefont{and}
  \bibinfo{author}{\bibfnamefont{E.}~\bibnamefont{Gheorghiu}},
  \bibinfo{journal}{Phys. Rev. E} \textbf{\bibinfo{volume}{81}},
  \bibinfo{pages}{021913} (\bibinfo{year}{2010}).

\bibitem[{\citenamefont{Sandu et~al.}(2011)\citenamefont{Sandu, Vrinceanu, and
  Gheorghiu}}]{Sandu2011}
\bibinfo{author}{\bibfnamefont{T.}~\bibnamefont{Sandu}},
  \bibinfo{author}{\bibfnamefont{D.}~\bibnamefont{Vrinceanu}},
  \bibnamefont{and}
  \bibinfo{author}{\bibfnamefont{E.}~\bibnamefont{Gheorghiu}},
  \bibinfo{journal}{Plasmonics} \textbf{\bibinfo{volume}{6}},
  \bibinfo{pages}{407} (\bibinfo{year}{2011}).

\bibitem[{\citenamefont{Sandu}(2013)}]{Sandu2013}
\bibinfo{author}{\bibfnamefont{T.}~\bibnamefont{Sandu}},
  \bibinfo{journal}{Plasmonics} \textbf{\bibinfo{volume}{8}},
  \bibinfo{pages}{391} (\bibinfo{year}{2013}).

\bibitem[{\citenamefont{Boyd}(2001)}]{Boyd2001}
\bibinfo{author}{\bibfnamefont{J.~P.} \bibnamefont{Boyd}},
  \emph{\bibinfo{title}{Chebyshev and Fourier Spectral Methods}}
  (\bibinfo{publisher}{Dover}, \bibinfo{address}{New York},
  \bibinfo{year}{2001}).

\bibitem[{\citenamefont{Chow and Yovanovich}(1982)}]{Chow1982}
\bibinfo{author}{\bibfnamefont{Y.~L.} \bibnamefont{Chow}} \bibnamefont{and}
  \bibinfo{author}{\bibfnamefont{M.~M.} \bibnamefont{Yovanovich}},
  \bibinfo{journal}{J. Appl. Phys.} \textbf{\bibinfo{volume}{53}},
  \bibinfo{pages}{8470} (\bibinfo{year}{1982}).

\bibitem[{\citenamefont{Landau and Lifshitz}(1984)}]{Landau1984}
\bibinfo{author}{\bibfnamefont{L.~D.} \bibnamefont{Landau}} \bibnamefont{and}
  \bibinfo{author}{\bibfnamefont{E.~M.} \bibnamefont{Lifshitz}},
  \emph{\bibinfo{title}{Electrodynamics of Continuous Media}}
  (\bibinfo{publisher}{Pergamon}, \bibinfo{address}{Oxford-New York},
  \bibinfo{year}{1984}).

\bibitem[{\citenamefont{Momoh et~al.}(2009)\citenamefont{Momoh, Sadiku, and
  Akujuobi}}]{Momoh2009}
\bibinfo{author}{\bibfnamefont{O.~D.} \bibnamefont{Momoh}},
  \bibinfo{author}{\bibfnamefont{M.~N.~O.} \bibnamefont{Sadiku}},
  \bibnamefont{and} \bibinfo{author}{\bibfnamefont{C.~M.}
  \bibnamefont{Akujuobi}}, \bibinfo{journal}{Microw. Opt. Tech. Lett.}
  \textbf{\bibinfo{volume}{51}}, \bibinfo{pages}{2361} (\bibinfo{year}{2009}).

\bibitem[{\citenamefont{Mansfield et~al.}(2001)\citenamefont{Mansfield,
  Douglas, and Garboczi}}]{Mansfield2001}
\bibinfo{author}{\bibfnamefont{M.~L.} \bibnamefont{Mansfield}},
  \bibinfo{author}{\bibfnamefont{J.~F.} \bibnamefont{Douglas}},
  \bibnamefont{and} \bibinfo{author}{\bibfnamefont{E.~J.}
  \bibnamefont{Garboczi}}, \bibinfo{journal}{Phys. Rev. E}
  \textbf{\bibinfo{volume}{64}}, \bibinfo{pages}{061401}
  (\bibinfo{year}{2001}).

\bibitem[{\citenamefont{Sandu}(2012)}]{Sandu2012}
\bibinfo{author}{\bibfnamefont{T.}~\bibnamefont{Sandu}}, \bibinfo{journal}{J.
  Nanopart. Res.} \textbf{\bibinfo{volume}{14}}, \bibinfo{pages}{905}
  (\bibinfo{year}{2012}).

\bibitem[{\citenamefont{Smythe}(1962)}]{Smythe1962}
\bibinfo{author}{\bibfnamefont{W.~R.} \bibnamefont{Smythe}},
  \bibinfo{journal}{J. Appl. Phys.} \textbf{\bibinfo{volume}{33}},
  \bibinfo{pages}{2966} (\bibinfo{year}{1962}).

\bibitem[{\citenamefont{Jackson}(2000)}]{Jackson2000}
\bibinfo{author}{\bibfnamefont{J.~D.} \bibnamefont{Jackson}},
  \bibinfo{journal}{Am. J. Phys.} \textbf{\bibinfo{volume}{68}},
  \bibinfo{pages}{789} (\bibinfo{year}{2000}).

\bibitem[{\citenamefont{Butler}(1980)}]{Butler1980}
\bibinfo{author}{\bibfnamefont{C.~M.} \bibnamefont{Butler}},
  \bibinfo{journal}{J. Appl. Phys.} \textbf{\bibinfo{volume}{51}},
  \bibinfo{pages}{5607} (\bibinfo{year}{1980}).

\bibitem[{\citenamefont{Maxwell}(1877)}]{Maxwell1877}
\bibinfo{author}{\bibfnamefont{J.~C.} \bibnamefont{Maxwell}},
  \bibinfo{journal}{Proc. London Math. Soc.} \textbf{\bibinfo{volume}{9}},
  \bibinfo{pages}{94} (\bibinfo{year}{1877}).

\bibitem[{\citenamefont{Jackson}(2002)}]{Jackson2002}
\bibinfo{author}{\bibfnamefont{J.~D.} \bibnamefont{Jackson}},
  \bibinfo{journal}{Am. J. Phys.} \textbf{\bibinfo{volume}{70}},
  \bibinfo{pages}{409} (\bibinfo{year}{2002}).

\bibitem[{\citenamefont{Ellenbogen et~al.}(2007)\citenamefont{Ellenbogen,
  Picconatto, and Burnim}}]{Ellenbogen2007}
\bibinfo{author}{\bibfnamefont{J.~C.} \bibnamefont{Ellenbogen}},
  \bibinfo{author}{\bibfnamefont{C.~A.} \bibnamefont{Picconatto}},
  \bibnamefont{and} \bibinfo{author}{\bibfnamefont{J.~S.}
  \bibnamefont{Burnim}}, \bibinfo{journal}{Phys. Rev. A}
  \textbf{\bibinfo{volume}{75}}, \bibinfo{pages}{042102}
  (\bibinfo{year}{2007}).

\bibitem[{\citenamefont{Lekner}(2011)}]{Lekner2011}
\bibinfo{author}{\bibfnamefont{J.}~\bibnamefont{Lekner}}, \bibinfo{journal}{J.
  Electrostat.} \textbf{\bibinfo{volume}{69}}, \bibinfo{pages}{11}
  (\bibinfo{year}{2011}).

\bibitem[{\citenamefont{Lekner}(2012{\natexlab{a}})}]{Lekner2012}
\bibinfo{author}{\bibfnamefont{J.}~\bibnamefont{Lekner}},
  \bibinfo{journal}{Proc. R. Soc. A} \textbf{\bibinfo{volume}{468}},
  \bibinfo{pages}{2433} (\bibinfo{year}{2012}{\natexlab{a}}).

\bibitem[{\citenamefont{Lekner}(2012{\natexlab{b}})}]{Lekner2012b}
\bibinfo{author}{\bibfnamefont{J.}~\bibnamefont{Lekner}}, \bibinfo{journal}{J.
  Appl. Phys.} \textbf{\bibinfo{volume}{111}}, \bibinfo{pages}{076102}
  (\bibinfo{year}{2012}{\natexlab{b}}).

\bibitem[{\citenamefont{Murovec and Brosseau}(2013)}]{Murovec2013}
\bibinfo{author}{\bibfnamefont{T.}~\bibnamefont{Murovec}} \bibnamefont{and}
  \bibinfo{author}{\bibfnamefont{C.}~\bibnamefont{Brosseau}},
  \bibinfo{journal}{Appl. Phys. Lett.} \textbf{\bibinfo{volume}{102}},
  \bibinfo{pages}{084105} (\bibinfo{year}{2013}).

\bibitem[{\citenamefont{Khair}(2013)}]{Khair2013}
\bibinfo{author}{\bibfnamefont{A.~S.} \bibnamefont{Khair}},
  \bibinfo{journal}{J. Appl. Phys.} \textbf{\bibinfo{volume}{114}},
  \bibinfo{pages}{134906} (\bibinfo{year}{2013}).

\bibitem[{\citenamefont{Prodan et~al.}(2008)\citenamefont{Prodan, Prodan, , and
  Miller}}]{Prodan2008}
\bibinfo{author}{\bibfnamefont{E.}~\bibnamefont{Prodan}},
  \bibinfo{author}{\bibfnamefont{C.}~\bibnamefont{Prodan}}, , \bibnamefont{and}
  \bibinfo{author}{\bibfnamefont{J.~H.} \bibnamefont{Miller}},
  \bibinfo{journal}{Biophys. J.} \textbf{\bibinfo{volume}{95}},
  \bibinfo{pages}{4174} (\bibinfo{year}{2008}).

\bibitem[{\citenamefont{Biasio et~al.}(2010)\citenamefont{Biasio, Ambrosone,
  and Cametti}}]{Biasio2010}
\bibinfo{author}{\bibfnamefont{A.~D.} \bibnamefont{Biasio}},
  \bibinfo{author}{\bibfnamefont{L.}~\bibnamefont{Ambrosone}},
  \bibnamefont{and} \bibinfo{author}{\bibfnamefont{C.}~\bibnamefont{Cametti}},
  \bibinfo{journal}{Biophys. J.} \textbf{\bibinfo{volume}{99}},
  \bibinfo{pages}{163} (\bibinfo{year}{2010}).

\bibitem[{\citenamefont{Stoy et~al.}(1982)\citenamefont{Stoy, Foster, and
  Schwan}}]{Stoy1982}
\bibinfo{author}{\bibfnamefont{R.}~\bibnamefont{Stoy}},
  \bibinfo{author}{\bibfnamefont{K.}~\bibnamefont{Foster}}, \bibnamefont{and}
  \bibinfo{author}{\bibfnamefont{H.}~\bibnamefont{Schwan}},
  \bibinfo{journal}{Phys. Med. Biol.} \textbf{\bibinfo{volume}{27}},
  \bibinfo{pages}{501} (\bibinfo{year}{1982}).

\bibitem[{\citenamefont{Lo et~al.}(1995)\citenamefont{Lo, Keese, and
  Giaever}}]{Lo1995}
\bibinfo{author}{\bibfnamefont{C.~M.} \bibnamefont{Lo}},
  \bibinfo{author}{\bibfnamefont{C.~R.} \bibnamefont{Keese}}, \bibnamefont{and}
  \bibinfo{author}{\bibfnamefont{I.}~\bibnamefont{Giaever}},
  \bibinfo{journal}{Biophys. J.} \textbf{\bibinfo{volume}{69}},
  \bibinfo{pages}{2800} (\bibinfo{year}{1995}).

\bibitem[{\citenamefont{Huang et~al.}(2010)\citenamefont{Huang, Sumpter,
  Meunier, Yushin, Portet, and Gogotsi}}]{Huang2010}
\bibinfo{author}{\bibfnamefont{J.}~\bibnamefont{Huang}},
  \bibinfo{author}{\bibfnamefont{B.~G.} \bibnamefont{Sumpter}},
  \bibinfo{author}{\bibfnamefont{V.}~\bibnamefont{Meunier}},
  \bibinfo{author}{\bibfnamefont{G.}~\bibnamefont{Yushin}},
  \bibinfo{author}{\bibfnamefont{C.}~\bibnamefont{Portet}}, \bibnamefont{and}
  \bibinfo{author}{\bibfnamefont{Y.}~\bibnamefont{Gogotsi}},
  \bibinfo{journal}{J. Mater. Res.} \textbf{\bibinfo{volume}{25}},
  \bibinfo{pages}{1525} (\bibinfo{year}{2010}).

\end{thebibliography}

\end{document}